\documentclass[twoside,12pt]{article}
\usepackage{epsfig}

\def\Journal#1#2#3#4{{#1} {#2} (#4) #3 }

\def\PLB{{\em Phys.\ Lett.} B}

\def\PRL{\em Phys.\ Rev.\ Lett.}
\def\PREV{\em Phys.\ Rev.}
\def\PREP{\em Phys.\ Rep.}

\def\PRD{{\em Phys.\ Rev.} D}
\def\PRC{{\em Phys.\ Rev.} C}

\newcommand{\be}{\begin{equation}}
\newcommand{\ee}{\end{equation}}
\newcommand{\bea}{\begin{eqnarray}}
\newcommand{\eea}{\end{eqnarray}}

\begin{document}

\title{ \vspace{1cm} From kinetic theory to dissipative fluid dynamics}
\author{B.\ Betz,$^1$ D.\ Henkel,$^1$ D.H.\ Rischke$^{1,2}$\\
\\
$^1$ Institut f\"ur Theoretische Physik, Johann Wolfgang 
Goethe-Universit\"at,\\
Max-von-Laue-Str.\ 1, D-60438 Frankfurt am Main, Germany\\
$^2$ Frankfurt Institute for Advanced Studies, \\
Ruth-Moufang-Str.\ 1, D-60438 Frankfurt am Main, Germany}
\maketitle
\begin{abstract} 
We present the results of deriving the Israel-Stewart equations of
relativistic dissipative fluid dynamics from kinetic
theory via Grad's 14-moment expansion. Working consistently
to second order in the Knudsen number, 
these equations contain several new terms
which are absent in previous treatments.
\end{abstract}
\section{Introduction}

Fluid dynamics has always been an
important qualitative tool to describe the collective
flow of hot and dense matter created in heavy-ion collisions
\cite{GreinerStoecker}. However, the great
success of the fluid dynamical model only came with nuclear collisions
at RHIC energies, where for the first time fluid dynamics was able
to describe flow observables, such as the elliptic flow
\cite{v2}, on a {\em quantitative\/} level \cite{KolbHeinz}.

The respective calculations \cite{KolbHeinz} 
of the elliptic flow at RHIC energies
were based on {\em ideal\/} fluid dynamics, 
i.e., all {\em dissipative\/}
effects were neglected. The good agreement between data and
the ideal fluid dynamical calculations gave rise to the notion that
``RHIC scientists serve up the perfect fluid''.
This implies that dissipative effects have to be small 
in order to not spoil the agreement with data.

However, the fluid dynamical equations of motion are partial differential
equations which require initial conditions in order to solve them
uniquely. In principle, these introduce infinitely many 
degrees of freedom that can be tuned to achieve
agreement with the data (once an appropriate ``freeze-out'' prescription
is adopted to convert the fluid into particles \cite{CooperFrye}). 
Since the initial conditions in heavy-ion collisions are not known
sufficiently precisely, one has to perform calculations within 
relativistic {\em dissipative\/} fluid dynamics and for
{\em various realistic\/} initial conditions
in order to confirm the smallness of dissipative effects.

Unfortunately, a consistent, stable, and causal formulation of 
relativistic dissipative fluid dynamics is far from trivial. The so-called
{\em first-order\/} theories due to Eckart \cite{Eckart} and Landau
\cite{Landau} have been shown \cite{Hiscock} to lead to unstable solutions and
to support acausal propagation of perturbations. A viable candidate
for a relativistic formulation of dissipative fluid dynamics, which
does not have these problems (for a large class of equations of state
\cite{Hiscock2}), is the so-called {\em second-order\/} theory 
due to Israel and Stewart \cite{IsraelStewart}.
In recent years, theoreticians have begun to apply this theory for
the description of collective flow in heavy-ion collisions
\cite{applications,Muronga,SongHeinz}.

In this paper we present the results of an analysis
(the details of which will be presented elsewhere \cite{BHR})
of deriving the Israel-Stewart 
equations of relativistic dissipative fluid dynamics from kinetic
theory, using the Boltzmann equation and Grad's 14-moment \cite{Grad}. We 
show that, working consistently to second order in the 
Knudsen number, additional terms arise in these 
equations that have been missed in the original paper 
\cite{IsraelStewart} and in subsequent treatments 
\cite{applications,Muronga,SongHeinz}.

This paper is organized as follows. After introducing our notation
in Sec.\ \ref{sec:prelim}, we discuss the power counting scheme in terms
of the Knudsen number (Sec.\ \ref{sec:powercount}). 
In Sec.\ \ref{sec:ISeqs} we present and discuss the Israel-Stewart equations.
A concluding section summarizes our results and gives an outlook to
future work.
Our units are $\hbar = c = k_B =1$, the metric tensor is
$g^{\mu \nu} = {\rm diag}(+,-,-,-)$. The scalar product of 4-vectors
$A^\mu$, $B^\mu$ is denoted as $A^\mu g_{\mu \nu} B^\nu = A^\mu B_\mu
\equiv A \cdot B$.

\section{Preliminaries} \label{sec:prelim}

The quantities that are evolved through the fluid dynamical equations
are the 4-current of net charge, $N^\mu$ (for the sake of simplicity,
we only consider a single conserved charge, the generalization to
several charges is straightforward \cite{PrakashVenugopalan}), 
and the symmetric rank-2 energy-momentum tensor, $T^{\mu \nu}$. 
In the following, we first discuss the tensor decomposition of 
these quantities and then give the corresponding
fluid dynamical equations which express the conservation of 
net charge and energy-momentum. Finally, the Navier-Stokes approximation
to relativistic dissipative fluid dynamics is discussed.

\subsection{\it Tensor Decomposition of $N^\mu$ and $T^{\mu \nu}$} 
\label{subsec:tensor}

The tensor decomposition of the net charge 4-current with
respective to an arbitrary 4-vector $u^\mu$ reads
\be
N^\mu = n\, u^\mu + \nu^\mu\;. \label{eq:nmu}
\ee
We shall identify $u^\mu$ with the {\em fluid 4-velocity\/} which is time-like
and normalized to one, $u \cdot u = 1$. This yields
$u^\mu = \gamma (1, {\bf v})$, where ${\bf v}$ is the fluid
3-velocity and $\gamma = (1-{\bf v}^2)^{-1/2}$ the corresponding Lorentz
gamma factor. Due to the normalization condition, the 4-vector $u^\mu$ has
only three independent components.
The component of $N^\mu$ in the direction of $u^\mu$ is 
the {\em net charge density\/} in the fluid rest frame,
$n \equiv N \cdot u$. The component orthogonal to $u^\mu$ is
the {\em diffusion current}, i.e., 
the flow of net charge relative to $u^\mu$, 
$\nu^\mu \equiv \Delta^{\mu \nu} N_\nu$.
Here, $\Delta^{\mu \nu} = g^{\mu \nu} - u^\mu u^\nu$ denotes
the projector onto the 3-space orthogonal to $u^\mu$. By construction,
$\Delta^{\mu \nu} u_\nu =0$, and thus also $\nu \cdot u = 0$. Therefore,
the 4-vector $\nu^\mu$ has only three independent components. 

The tensor decomposition of the energy-momentum tensor reads
\be
T^{\mu \nu} = \epsilon\, u^\mu u^\nu - (p+\Pi)\,
\Delta^{\mu \nu} + 2\, q^{(\mu} u^{\nu)} +\pi^{\mu \nu}\;, \label{eq:tmunu}
\ee
where $\epsilon \equiv u^\mu T_{\mu \nu} u^\nu$ is the {\em energy density\/}
in the fluid rest frame. The projection $p+\Pi \equiv
-\frac{1}{3} \, \Delta^{\mu \nu} T_{\mu \nu}$ is the sum
of {\em  thermodynamic pressure}, $p$, and {\em bulk viscous pressure}, $\Pi$.
The {\em heat flux current}, i.e., the flow of energy relative to $u^\mu$, is
$q^\mu \equiv \Delta^{\mu \nu} T_{\nu \lambda} u^\lambda$. By construction,
$q \cdot u = 0$, and $q^\mu$ has only three independent components. 
The notation $a^{(\alpha_1 \cdots \alpha_n)}$
stands for symmetrization in all Lorentz indices, e.g.,
$a^{(\mu \nu)} \equiv \frac{1}{2} \left(a^{\mu \nu} + a^{\nu \mu}\right)$.
Finally, $\pi^{\mu\nu} \equiv T^{<\mu \nu>}$ is the {\em shear
stress tensor}, where $a^{< \mu \nu >} \equiv \left( 
\Delta_\alpha^{\hspace*{0.2cm}(\mu} 
\Delta^{\nu)}_{\hspace*{0.3cm}\beta} -
\frac{1}{3}\, \Delta^{\mu \nu} \Delta_{\alpha \beta} \right)
a^{\alpha \beta}$ denotes the symmetrized, traceless spatial projection.
Thus, by construction, $\pi^{\mu\nu} u_\mu 
= \pi^{\mu \nu} u_\nu = \pi^{\mu}_{\hspace*{0.2cm}\mu} =0$.
This implies that $\pi^{\mu \nu}$ has only five independent components.

\subsection{\it Fluid Dynamical Equations of Motion} \label{subsec:eom}

The conservation of net charge reads
\be \label{eq:chargecons}
\partial \cdot N = \dot{n} + n\, \theta + \partial \cdot \nu = 0\;, 
\ee
where $\dot{a} \equiv u \cdot \partial \, a$ denotes the
comoving derivative. In the fluid rest frame, $u^\mu_{\rm RF} =
(1,0,0,0)$, it is simply the time derivative, $\dot{a}_{\rm RF} =
\partial_t a$. The quantity $\theta \equiv \partial \cdot u$ is
the so-called {\em expansion scalar}.

The equation for energy-momentum conservation, 
$\partial_\mu T^{\mu \nu} =0$ is a 4-vector and can also be tensor-decomposed
into a component parallel to $u^\mu$,
\be \label{eq:encons}
u_\nu \, \partial_\mu T^{\mu \nu}
= \dot{\epsilon} + (\epsilon +p+\Pi)\, \theta + \partial \cdot q
- q\cdot \dot{u}- \pi^{\mu \nu}\, \partial_\mu u_\nu =0\;,
\ee
which represents the {\em conservation of energy}, and into the
three independent components
orthogonal to $u^\mu$, $\Delta^{\mu \nu} \, \partial^\lambda T_{\nu \lambda} 
=0$, which can be cast into the form
\be \label{eq:acc}
(\epsilon +p )\, \dot{u}^\mu = \nabla^{\mu} (p + \Pi) 
- \Pi\, \dot{u}^\mu
- \Delta^{\mu \nu} \dot{q}_\nu - q^\mu \, \theta - q \cdot \partial \, u^\mu
- \Delta^{\mu \nu} \,\partial^\lambda \pi_{\nu \lambda}\;,
\ee
where the notation $\nabla^{\mu} \equiv \Delta^{\mu \nu} \partial_\nu$ stands
for the 3-gradient, i.e., the spatial gradient in the fluid rest frame.
Equation (\ref{eq:acc}) is the so-called {\em acceleration equation}. 
The first term on the r.h.s.\ expresses the fact that
changes in the fluid 4-velocity are driven by pressure gradients.

The problem with the five equations of motion (\ref{eq:chargecons}),
(\ref{eq:encons}), and (\ref{eq:acc}) is that, for any given
fluid 4-velocity $u^\mu$, they contain 15 unknowns, 
$\epsilon,\, p,\, n,\, \Pi$, the three components of $\nu^\mu$, the
three components of $q^\mu$, and the five components 
of $\pi^{\mu \nu}$. The choice of reference frame does not alleviate this
problem: in the Eckart frame $\nu^\mu \equiv 0$, and in the
Landau frame $q^\mu \equiv 0$, however, then $u^\mu$ is no longer fixed, but
becomes a dynamical quantity, and one still has 15 unknowns. In principle, 
there are two strategies to continue. 
\begin{enumerate}
\item One assumes the fluid to be
in {\em local thermodynamical equilibrium}. Then, all dissipative
quantities vanish, $\Pi = q^\mu$ (or $\nu^\mu$) $= \pi^{\mu \nu}\equiv 0$, 
which eliminates nine out of the 15 unknowns. This is
the so-called {\em ideal-fluid limit}. In this case, the reference frame
is uniquely defined. The sixth
remaining unknown, for instance $p$, is determined by choosing an
equation of state for the fluid in the form $p=p(\epsilon, n)$. Then,
the system of five equations (\ref{eq:chargecons}), (\ref{eq:encons}),
and (\ref{eq:acc}) can be uniquely solved for $\epsilon,\, n$, and $u^\mu$.
\item One provides equations that determine the dissipative
quantities: the equations of {\em dissipative fluid dynamics}.
One distinguishes {\em first-order\/} and {\em second-order theories\/} 
of dissipative fluid dynamics. A first-order theory is, e.g., the
{\em Navier-Stokes\/} (NS) approximation where the dissipative quantities
$\Pi,\, q^\mu$ (or $\nu^\mu$), and 
$\pi^{\mu \nu}$ are expressed solely in terms of the
primary variables $\epsilon, \, p, \, n,\, u^\mu$, or gradients thereof.
A second-order theory is, for instance, the so-called {\em Israel-Stewart\/}
(IS) theory \cite{IsraelStewart}. Here, the dissipative quantities
are independent dynamical quantities whose evolution is governed by
differential equations similar to the fluid dynamical equations
(\ref{eq:chargecons}), (\ref{eq:encons}), and (\ref{eq:acc}).
\end{enumerate}

\subsection{\it Navier-Stokes Approximation} \label{subsec:nsapp}

In the NS approximation, the dissipative quantities
$\Pi,\, q^\mu,\, \pi^{\mu \nu}$ read:
\bea
\Pi_{\rm NS} & = & - \zeta\, \theta\;, \label{eq:PiNS} \\
q^\mu_{\rm NS} & = & \frac{\kappa}{\beta}\, 
\frac{n}{\beta (\epsilon +p)}\, \nabla^\mu \alpha\;, \label{eq:qNS}\\
\pi^{\mu \nu}_{\rm NS} & = &
2 \, \eta \, \sigma^{\mu \nu}\; , \label{eq:piNS}
\eea
where $\beta \equiv 1/T$ and $\alpha \equiv \beta \mu$; $\mu$ is the
{\em chemical potential\/} associated with the net charge density $n$. The
quantitites $\zeta,\, \kappa$, and $\eta$ are the {\em bulk viscosity},
{\em thermal conductivity}, and {\em shear viscosity\/} coefficients.
The {\em shear tensor\/} is defined as $\sigma^{\mu \nu} \equiv
\nabla^{< \mu} u^{\nu >}$. Note that Eq.\ (\ref{eq:qNS}) holds in
the Eckart frame; in the Landau frame, simply
replace $q^\mu \rightarrow - \nu^\mu (\epsilon +p)/n$.

Since $\Pi,\, q^\mu$ (or $\nu^\mu$), and $\pi^{\mu \nu}$ 
are solely given in terms of the primary variables, 
one can simply insert the expressions (\ref{eq:PiNS}),
(\ref{eq:qNS}), and (\ref{eq:piNS}) into the Eqs.\ (\ref{eq:chargecons}),
(\ref{eq:encons}), and (\ref{eq:acc}) and obtain a closed set of equations
of motion: the relativistic generalization of the non-relativistic NS
equations. As mentioned in the Introduction, the problem with these 
equations is that they lead to unstable solutions and 
support acausal propagation of perturbations.

\section{Power counting} \label{sec:powercount}

In order to derive the IS equations, 
we first have to discuss the length scales appearing in
fluid dynamics. Then we identify the Knudsen number as the small
quantity in terms of which one can do consistent power counting.
The IS equations will then emerge at second order in an expansion
in powers of the Knudsen number.

\subsection{\it Scales in fluid dynamics} \label{subsec:scales}

In principle, there are three length scales in fluid dynamics, two
microscopic scales and one macroscopic scale. The two microscopic scales
are the {\em thermal wavelength}, $\lambda_{\rm th} \sim \beta$, and the
{\em mean free path}, $\ell_{\rm mfp} \sim (\langle \sigma \rangle n)^{-1}$,
where $\langle \sigma \rangle$ is the average cross section. The macroscopic
scale, $L_{\rm hydro}$, is the scale over which the fluid 
fields $\epsilon,\, n, u^\mu, \ldots$ vary, i.e., 
gradients of these fields are typically
of order $\partial_\mu \sim L_{\rm hydro}^{-1}$. Note that, since
$n^{-1/3} \sim \beta \sim \lambda_{\rm th}$, the
thermal wavelength can be interpreted as the
interparticle distance. Note also that the notion
of a mean free path requires the existence of well-defined quasi-particles.
In strongly coupled theories, the quasi-particle concept is no longer 
valid. In this case, there are only two scales, $\lambda_{\rm th}$ and
$L_{\rm hydro}$.

As a first important result let us note that the ratios of the transport
coefficients $\zeta,\, \kappa/\beta$, and $\eta$ to the
entropy density are solely determined by the ratio of
the two microscopic length scales, $\ell_{\rm mfp}/\lambda_{\rm th}$. 
We demonstrate this explicitly for the shear viscosity to entropy density
ratio, similar arguments also hold for the other transport coefficients.
For the proof, note that $\eta \sim (\langle \sigma \rangle
\lambda_{\rm th})^{-1}$ and $n \sim T^3 \sim s$. Then,
\be \label{eq:etas}
\frac{\ell_{\rm mfp}}{\lambda_{\rm th}}
\sim \frac{1}{\langle \sigma \rangle n}\, \frac{1}{\lambda_{\rm th}}
\sim \frac{1}{\langle \sigma \rangle \lambda_{\rm th}}\,
\frac{1}{n} \sim \frac{\eta}{s}\;.
\ee
There are two limiting cases, (a) the {\em dilute-gas limit},
$\ell_{\rm mfp}/\lambda_{\rm th}\sim \eta/s \rightarrow \infty$, and (b) the
{\em ideal-fluid limit}, $ \ell_{\rm mfp}/\lambda_{\rm th}\sim \eta/s 
\rightarrow 0$. Estimating $\ell_{\rm mfp} \sim \langle \sigma \rangle^{-1}
\lambda_{\rm th}^3$, the first case corresponds to
$\langle \sigma \rangle /\lambda_{\rm th}^2 \rightarrow 0$, i.e.,
the interaction cross section is much smaller than the area given by
the thermal wavelength. In other words, the average distance between
collisions is much larger than the interparticle distance. In this
sense, the dilute-gas limit can be interpreted as a {\em weak-coupling limit}.
Similarly, the ideal-fluid limit corresponds to $\langle \sigma \rangle
/\lambda_{\rm th}^2 \rightarrow \infty$. This is the somewhat academic
case when interactions happen on a scale much smaller than the interparticle
distance. In this sense, this is the limit of {\em infinite coupling}, i.e.,
the interactions are so strong that the fluid assumes
locally and instantaneously a state of thermodynamical equilibrium.

For {\em any\/} value of $\eta/s$ (and, analogously,
$\zeta/s$ and $\kappa/(\beta s)$) between these two limits, the
equations of dissipative fluid dynamics may be applied for the 
description of the system. The situation is particularly interesting for
$\ell_{\rm mfp}/\lambda_{\rm th} \sim 
\eta/s \sim 1$ or, equivalently, $\langle \sigma \rangle \sim 
\lambda_{\rm th}^2 \sim T^{-2}$. 
In this case, there is only a {\em single\/} microscopic scale 
$\lambda_{\rm th}$ in the problem. This occurs, for instance, in
strongly coupled theories without well-defined quasi-particles.

One may derive the equations of dissipative fluid dynamics
in terms of a gradient expansion.
However, in order to be able to truncate this expansion 
after a finite number of terms one has to require that gradients are small
or, equivalently, that the {\em Knudsen number\/} is small. This is
discussed in the following subsection.

\subsection{\it Expansion in Terms of the Knudsen Number}
\label{subsec:knudsen}

The Knudsen number is defined as
$K \equiv \ell_{\rm mfp}/L_{\rm hydro}$. Since $L_{\rm hydro}^{-1} \sim
\partial_\mu$, an expansion in
terms of $K$ is equivalent to a gradient expansion, i.e., an expansion
in terms of powers of $\ell_{\rm mfp} \, \partial_\mu$.

We can now establish a second important result: provided that the
dissipative quantities $\Pi,\, q^\mu$ (or $\nu^\mu$), and $\pi^{\mu \nu}$ 
do not differ too much from their NS values, the ratios of these
quantities to the energy density are proportional to the
Knudsen number. We demonstrate this explicitly for the bulk viscous pressure;
similar arguments apply for the heat flux current and the shear stress
tensor. We use the fundamental relation of thermodynamics,
$\epsilon + p = Ts + \mu n$, to estimate $\beta\, \epsilon \sim 
\lambda_{\rm th}\, \epsilon \sim s$ and we employ Eq.\ (\ref{eq:PiNS}) to
write
\be \label{eq:Piepsilon}
\frac{\Pi}{\epsilon} \sim
\frac{\Pi_{\rm NS}}{\epsilon} \sim
\frac{\zeta \, \theta}{\epsilon} \sim 
\frac{\zeta}{\lambda_{\rm th}\, \epsilon}\, \lambda_{\rm th}\, \theta \sim
\frac{\zeta}{s} \, 
\frac{\lambda_{\rm th}}{\ell_{\rm mfp}}\, \ell_{\rm mfp}\, \partial_\mu
u^\mu \sim \frac{\zeta}{s}\, \left( \frac{\ell_{\rm mfp}}{\lambda_{\rm th}}
\right)^{-1} \, K \, u^\mu \sim K \;.
\ee
In the last step, we have employed Eq.\ (\ref{eq:etas}) and the fact that
$u^\mu \sim 1$. The result is remarkable in the sense that $\Pi/\epsilon$
is {\em only\/} proportional to $K$, and {\em independent\/} of the ratio of 
viscosity to entropy density. The reason is
that this ratio drops out on account of Eq.\ (\ref{eq:etas}). 

As a corollary to Eq.\ (\ref{eq:Piepsilon}), we conclude that if
the Knudsen number is small, $K \sim \delta \ll 1$,
the dissipative quantities are small compared to the primary variables.
The system is close to local thermodynamical equilibrium or, in other
words, close to the ideal-fluid limit. The equations
of dissipative fluid dynamics can then be systematically 
and in a well-controlled manner derived in terms
of a gradient expansion or, equivalently, in terms of
a power series in $K$ or, equivalently because of Eq.\ 
(\ref{eq:Piepsilon}), in terms of
powers of dissipative quantities. At zeroth order in $K$, one obtains
the equations of ideal fluid dynamics. At first order in $K$, one
obtains the NS equations. At second order in $K$, the IS equations
emerge. 

Finally note that the independence of the ratio of
dissipative quantities to primary variables from
the viscosity to entropy density ratio has 
important phenomenological consequences.
It guarantees that, provided that gradients of
the macroscopic fluid fields (and, thus, $K$)
are sufficiently small, the NS equations are still 
valid and applicable for the description of
systems with large $\eta/s$, e.g.\  
water at room temperate and atmospheric pressure.

\section{The Israel-Stewart Equations} \label{sec:ISeqs}

The IS equations of relativistic dissipative fluid dynamics
can be derived from the Boltzmann equation  
via Grad's 14-moment method \cite{Grad}. The details of this
derivation will be presented elsewhere \cite{BHR}. 
To second order in dissipative quantities
(or equivalently, because of 
Eq.\ (\ref{eq:Piepsilon}), to second order in $K$) the equations read:
\bea
\Pi & = & \Pi_{\rm NS} - \tau_\Pi\, \dot{\Pi} \nonumber \\
& + &  \tau_{\Pi q}\, q \cdot \dot{u}
 - \ell_{\Pi q}\, \partial \cdot q
 - \zeta \, \hat{\delta}_0\, \Pi\, \theta \nonumber \\
& + &  \lambda_{\Pi q} \, q \cdot \nabla \alpha
 +  \lambda_{\Pi \pi} \, \pi^{\mu \nu} \sigma_{\mu \nu}\;, 
\label{eq:PiIS} \\
q^\mu  & = & q^\mu_{\rm NS}  - \tau_q \, \Delta^{\mu \nu}\dot{q}_\nu 
\nonumber \\
& - &  \tau_{q \Pi}\, \Pi\, \dot{u}^\mu 
- \tau_{q \pi}\, \pi^{\mu \nu}\, \dot{u}_\nu \nonumber 
 +    \ell_{q \Pi}\, \nabla^{\mu} \Pi  - 
\ell_{q \pi}\, \Delta^{\mu \nu}\, \partial^\lambda 
\pi_{\nu \lambda} +  \tau_q \, \omega^{\mu \nu} \, q_\nu
 - \frac{\kappa}{\beta}\, \hat{\delta}_1\, q^\mu \, \theta \nonumber \\
& - &  \lambda_{qq}\, \sigma^{\mu \nu}\, q_\nu 
+  \lambda_{q \Pi}\, \Pi \, \nabla^\mu \alpha
+  \lambda_{q \pi}\, \pi^{\mu \nu}\, \nabla_\nu \alpha\; ,
\label{eq:qIS}\\
\pi^{\mu \nu} & = &
\pi^{\mu \nu}_{\rm NS}  -
\tau_\pi\, \dot{\pi}^{<\mu \nu>}  \nonumber \\
& + &   2\, \tau_{\pi q}\, q^{<\mu} \dot{u}^{\nu>}
 +  2 \, \ell_{\pi q}\, \nabla^{<\mu} q^{\nu>}
 +  2\, \tau_\pi\, \pi_\lambda^{\hspace*{0.1cm}<\mu} 
\omega^{\nu> \lambda}
 -  2\, \eta\, \hat{\delta}_2\, \pi^{\mu \nu}\, \theta \nonumber \\
& - &    2\, \tau_\pi \, \pi_\lambda^{\hspace*{0.1cm}<\mu} 
\sigma^{\nu> \lambda}-  2\, \lambda_{\pi q}\, q^{<\mu} 
\nabla^{\nu>} \alpha + 2\, \lambda_{\pi \Pi}\, \Pi\, \sigma^{\mu \nu}\;,
\label{eq:piIS}
\eea
where $\omega^{\mu \nu} \equiv \frac{1}{2} \, 
\Delta^{\mu \alpha} \Delta^{\nu \beta} \left( \partial_\alpha u_\beta
- \partial_\beta u_\alpha\right)$ is the {\em vorticity}.
We now discuss these equations.
\begin{enumerate}
\item The transport coefficients $\zeta, \, \kappa, \, \eta$,
the relaxation times $\tau_\Pi,\, \tau_q,\, \tau_\pi$, the
coefficients $\tau_{\Pi q},\, \tau_{q \Pi} , \, \tau_{q \pi}$, 
$\tau_{\pi q},\, \ell_{\Pi q}, \, \ell_{q \Pi}, \,
\ell_{q \pi},\, \ell_{\pi q},\, \lambda_{\Pi q},\,  
\lambda_{\Pi \pi}, \, \lambda_{qq}, \, \lambda_{q \Pi}, \, 
\lambda_{q \pi},\, \lambda_{\pi q}, \, \lambda_{\pi \Pi}$
are (complicated) functions of $\alpha,\, \beta$, divided 
by tensor coefficients of the second moment of the
collision integral, for details, see Ref.\ \cite{BHR}.
As one sends the cross section in the
collision integral to infinity, all these coefficients go to zero.
Since then also the NS values vanish, one ends
up with the trivial solution
$\Pi = q^\mu = \pi^{\mu \nu} \equiv 0$ to the IS equations, i.e., one
recovers the ideal-fluid limit. This is consistent with the discussion
in Sec.\ \ref{subsec:scales}.
\item The coefficients $\hat{\delta}_0,\,
\hat{\delta}_1,\, \hat{\delta}_2$ are (complicated) functions of
$\alpha,\, \beta$.
\item The form of the equations is invariant of the calculational frame
(Eckart, Landau, \ldots), however, the values of the coefficients are
frame-dependent. The obvious reason is that the physical interpretation
of the dissipative quantities is frame-dependent. For instance, in the
Eckart frame, $q^\mu$ is the heat flux current, while in the
Landau frame, $q^\mu \equiv - \nu^\mu (\epsilon +p)/n$ is the (negative of
the) diffusion current, multiplied by the specific enthalpy.
Details are given in Ref.\ \cite{BHR}.
\item The NS terms (the first terms on the r.h.s.) are of {\em first order\/}
in $K$, all other terms are of {\em second order\/} in $K$. Consequently,
dropping the latter one obtains the NS equations 
(\ref{eq:PiNS}), (\ref{eq:qNS}), and (\ref{eq:piNS}).
\item The so-called {\em simplified\/} IS equations (in the terminology
of Ref.\ \cite{SongHeinz}) emerge by keeping only the first lines
of Eqs.\ (\ref{eq:PiIS}), (\ref{eq:qIS}), and (\ref{eq:piIS}).
The resulting equations have the simple interpretation that
the dissipative quantities $\Pi,\, q^\mu$, and $\pi^{\mu \nu}$
{\em relax\/} to their corresponding NS values on time scales
$\tau_\Pi,\, \tau_q$, and $\tau_\pi$, respectively.
\item Considering the {\em full\/} IS equations (\ref{eq:PiIS}),
(\ref{eq:qIS}), and (\ref{eq:piIS}), 
for times $t<\tau_i$, $i=\Pi,q,\pi$, the dissipative quantities $\Pi,\, 
q^\mu,\, \pi^{\mu \nu}$ are driven towards their NS values. 
Once they are reasonably close to these, the first terms
on the r.h.s.\ largely cancel against the l.h.s.. The further evolution,
for times $t > \tau_i$, is then determined by the remaining, second-order
terms. These terms therefore constitute important corrections
for times $t> \tau_i$ and should not be neglected.
\item The first two terms in the second line of Eq.\ (\ref{eq:PiIS}),
the first five terms in the second line of Eq.\ (\ref{eq:qIS}), and 
the first three terms in the second line of Eq.\ (\ref{eq:piIS})
were also obtained by Israel and Stewart \cite{IsraelStewart}, while the
remaining second-order terms were missed or neglected. Presumably,
Israel and Stewart made the assumption that second-order terms containing
$\theta$, $\sigma^{\mu \nu}$, or $\nabla^\mu \alpha$ 
are even smaller than suggested by power counting in terms of $K$.
The last two terms in the second line of Eq.\ (\ref{eq:PiIS}), 
the last four terms in the second line of Eq.\ (\ref{eq:qIS}), and 
the last three terms in the second line of Eq.\ (\ref{eq:piIS})
were also obtained by Muronga \cite{Muronga}, while the other second-order
terms do not appear in that paper. A possible reason is that the corresponding
treatment is based on the phenomenological approach to derive the
IS equations and terms that do not generate entropy are absent.
The terms in the third line of Eqs.\ (\ref{eq:PiIS}), (\ref{eq:qIS}),
and (\ref{eq:piIS}) were neither given by Israel and Stewart 
\cite{IsraelStewart} nor by Muronga \cite{Muronga} and are thus genuinely
new to this paper (with one exception discussed below).
\item If we set $\Pi = q^\mu=0$ in Eq.\ (\ref{eq:piIS}), the resulting
equation for $\pi^{\mu \nu}$ is identical to that found in Ref.\ \cite{Baier}.
In particular, the first term in the third line was
already obtained in that paper, where it appeared in the form
$(\lambda_1/\eta^2)\, \pi_\lambda^{\hspace*{0.1cm}<\mu} 
\pi^{\nu> \lambda}$. Using the NS value (\ref{eq:piNS})
for $\pi^{\nu \lambda}$, which is admissible because we are computing
to second order in $K$, to this order this is identical to
$2 (\lambda_1/\eta)\,\pi_\lambda^{\hspace*{0.1cm}<\mu} 
\sigma^{\nu> \lambda}$. By comparison with Eq.\ (\ref{eq:piIS}), 
we thus get a prediction for the coefficient
$\lambda_1$ from kinetic theory, $\lambda_1 \equiv \tau_\pi\, \eta$,
in agreement with Ref.\ \cite{Baier}. Note, however, that this
discussion so far neglects additional terms which arise at second order
in $K$ when expanding the second moment of the collision integral. (This
was already noted in Ref.\ \cite{Baier}.) This will
change the coefficient of the respective term such that it is no longer
equal to $\tau_\pi$. It will therefore also lead to a different result for
$\lambda_1$. In a recent study \cite{Moore} a complete calculation was
performed. 
\end{enumerate}

\section{Conclusion}

In this paper, we have discussed the full Israel-Stewart (IS) 
equations of relativistic dissipative fluid dynamics as they
emerge from applying Grad's 14-moment expansion to
the Boltzmann equation and truncating dissipative effects
at second order in the Knudsen number $K= \ell_{\rm mfp}/L_{\rm hydro}$. 
Our treatment is not restricted to the shear stress tensor, but also
contains the bulk viscous pressure and the heat flux current. It is thus
also applicable to non-conformal systems with non-vanishing 
net charge density.

We have shown that, in comparison to previous discussions
\cite{IsraelStewart,applications,Muronga}, 
additional second-order terms appear. 
One of these terms was found already in Ref.\ \cite{Baier}.
The details of the derivation of the full second-order IS equations
will be presented elsewhere \cite{BHR}.
The second-order terms are multiplied with coefficients whose values
depend on the calculational frame. Explicit expressions will be reported
elsewhere \cite{BHR}.
Future directions of work comprise the generalization to a system
of various particle species \cite{PrakashVenugopalan}, as well as
the numerical implementation and application to modelling the dynamics
of heavy-ion collisions.
\\[0.2cm] 
{\em Acknowledgment.}
We would like to thank G.\ Moore, U.W.\ Heinz, P.\ Romatschke, 
G.\ Torrieri, and U.A.\ Wiedemann for enlightening discussions.
The work of B.B.\ was supported by the Helmholtz Research School H-QM.

\end{document}